\documentclass[12pt]{article}
\usepackage{graphicx}
\usepackage{color}
\newcommand{\beq}{\begin{equation}}
\newcommand{\eeq}{\end{equation}}
\newcommand{\bea}{\begin{eqnarray}}
\newcommand{\eea}{\end{eqnarray}}

\newcommand{\gsim}{\lower.7ex\hbox{$
\;\stackrel{\textstyle>}{\sim}\;$}}
\newcommand{\lsim}{\lower.7ex\hbox{$
\;\stackrel{\textstyle<}{\sim}\;$}}

\newcommand{\eod}{\end{document}}

\usepackage{epsfig}

\definecolor{verm}{rgb}{0.8,0.1,0.0}

\begin{document}
\thispagestyle{empty}
\vspace*{-22mm}

\begin{flushright}

UND-HEP-18-BIG\hspace*{.08em}01[V3] \\

\end{flushright}

\vspace*{0.7mm}

\begin{center}
{\Large {\bf The Dynamics of Beauty \& Charm Hadrons and top quarks 
in the Era of the LHCb \& Belle II and ATLAS/CMS -- \\ 
Non-perturbative QCD \& Many-body Final States 
\footnote{Invited talk given at the {\em 2018 Epiphany Conference}}}}

\vspace*{7mm}

{\bf I.I.~Bigi$^a$} \\
\vspace{3mm}
$^a$ {\sl Department of Physics, University of Notre Dame du Lac}\\
{\sl Notre Dame, IN 46556, USA} 

\vspace*{-.8mm}

{\sl email addresses: ibigi@nd.edu} 

\vspace*{9mm}

\begin{center}
{\Large {\em This article is dedicated to Timothy O'Meara, mathematician \& first lay Provost of the University of Notre Dame du Lac.}}
\end{center}

\vspace*{9mm}

{\bf Abstract}\vspace*{-1.5mm}\\
\end{center}

\noindent
Our community has to apply {\em non}-perturbative QCD on different levels of flavor dynamics 
in strange, charm \& beauty hadrons and even for top quarks. We need {\em consistent} parameterization of the CKM matrix and describe  
weak decays of beauty hadrons with {\em many-body} final states. 
It is crucial to use the {\em Wilsonian} OPE as much as possible and discuss "duality" in the worlds of quarks and hadrons. The pole mass of heavy quarks is  
{\em not} well-defined on the {\em non}-perturbative level -- i.e., it is {\em not} Borel summable in total QCD. 
We need a novel team to combine the strengths of our tools from MEP and HEP.

\vspace{3mm}

\hrule

\tableofcontents

\vspace{5mm}

\hrule\vspace{5mm}


\section{Prologue}
\label{PROLOGUE}
\begin{figure}[h!]
\begin{center}
\includegraphics[width=5cm]{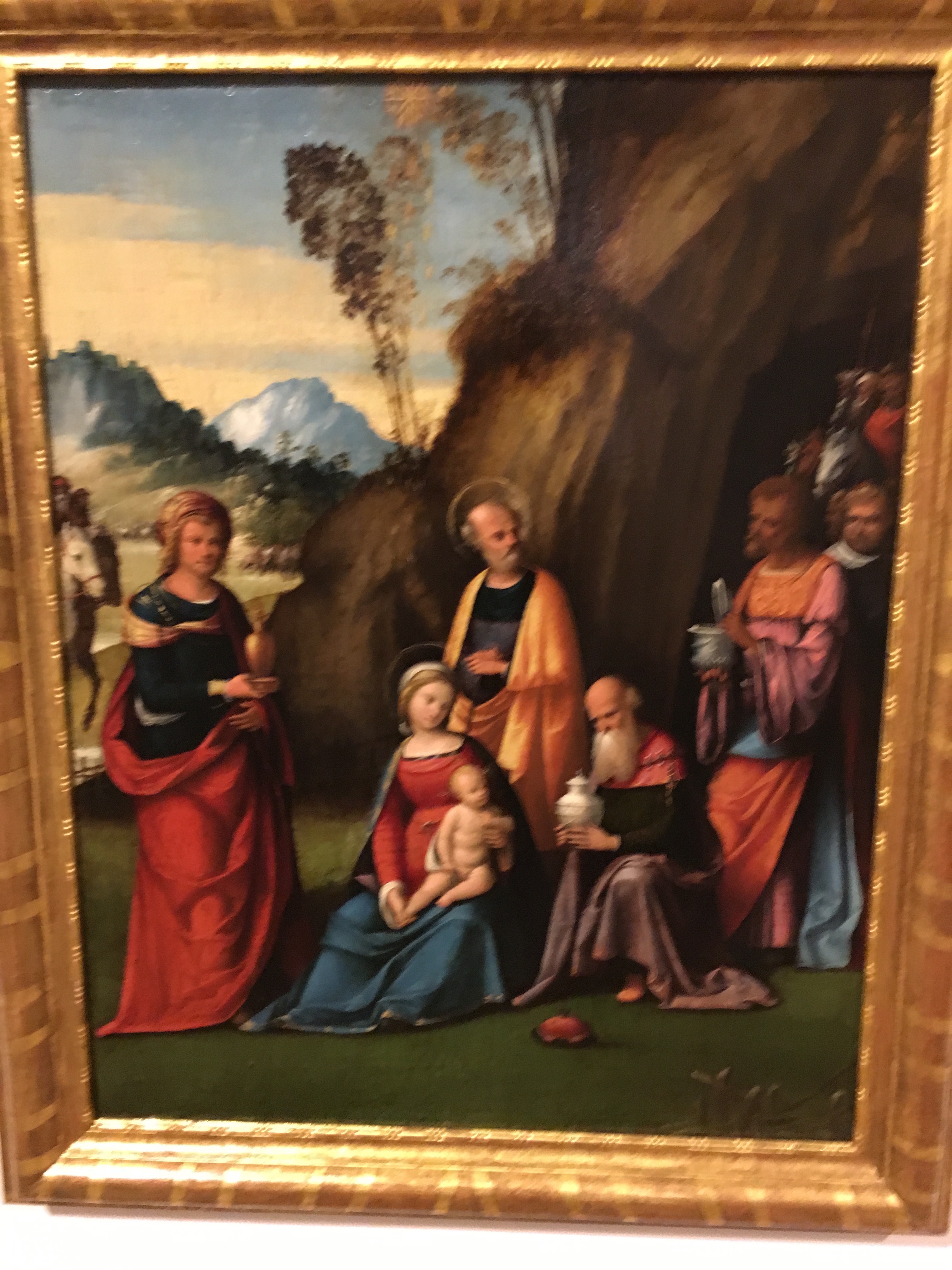}
\end{center}
 \vspace{-1cm}
\caption{Painting of `Epiphany' in a museum in Krakow (picture taken by IIB)}
\label{Museum1}
\end{figure}
I have truly enjoyed the 2018 Epiphany Conference in Krakow, learnt about fundamental dynamics -- and 
the `landscapes' of history \& art on the true European scale. Very special event happened on January 6 long time ago  
(see the {\bf Figure \ref{Museum1}}): three sages came to meet with the Christ child. 

The old center of the city of Krakow is an amazing part of the European culture. I try to show that by pictures I took on 
Jan. 13, 2018: 
\begin{itemize}
\item 
The Barbican in Krakow is just outside of one of the gates on the north wall and very close to the Guesthouse. The 
Barbican was to protect the city against the `barbarians', see the {\bf Figure \ref{KrakowBarbican}}.  
\begin{figure}[h!]
\begin{center}
\includegraphics[scale=0.30]{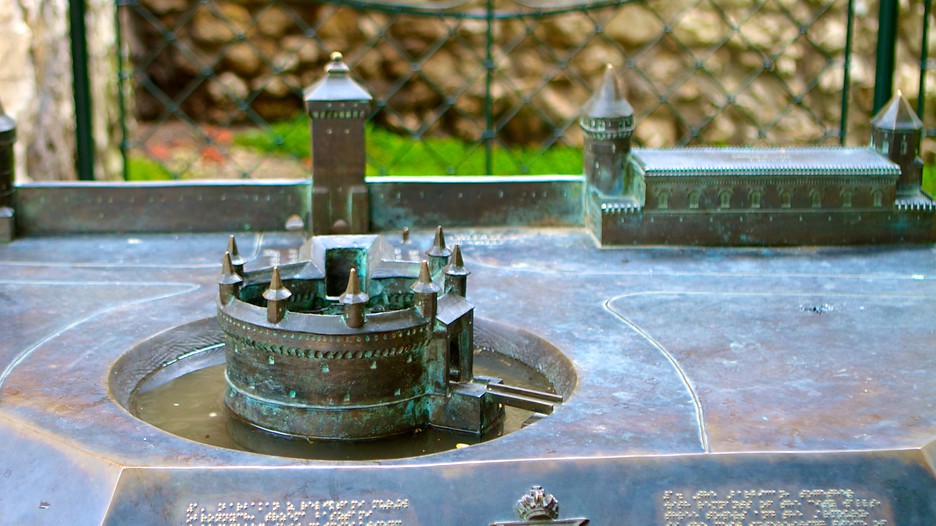}
\end{center}
\vspace{-0.5cm}
\caption{Barbican in Krakow (Model)} 
\label{KrakowBarbican}
\end{figure}
An analogy to use Dalitz plots to probe the impact of New Dynamics (ND): `barbarians' = `perturbative QCD'?

\item 
Copericus was a student in the Jagiellonian University of Krakow that had large impact on the understand of our Universe then; 
see the {\bf Figure \ref{COPERICUS}}.
\begin{figure}[h!]
\begin{center}
\includegraphics[scale=0.05]{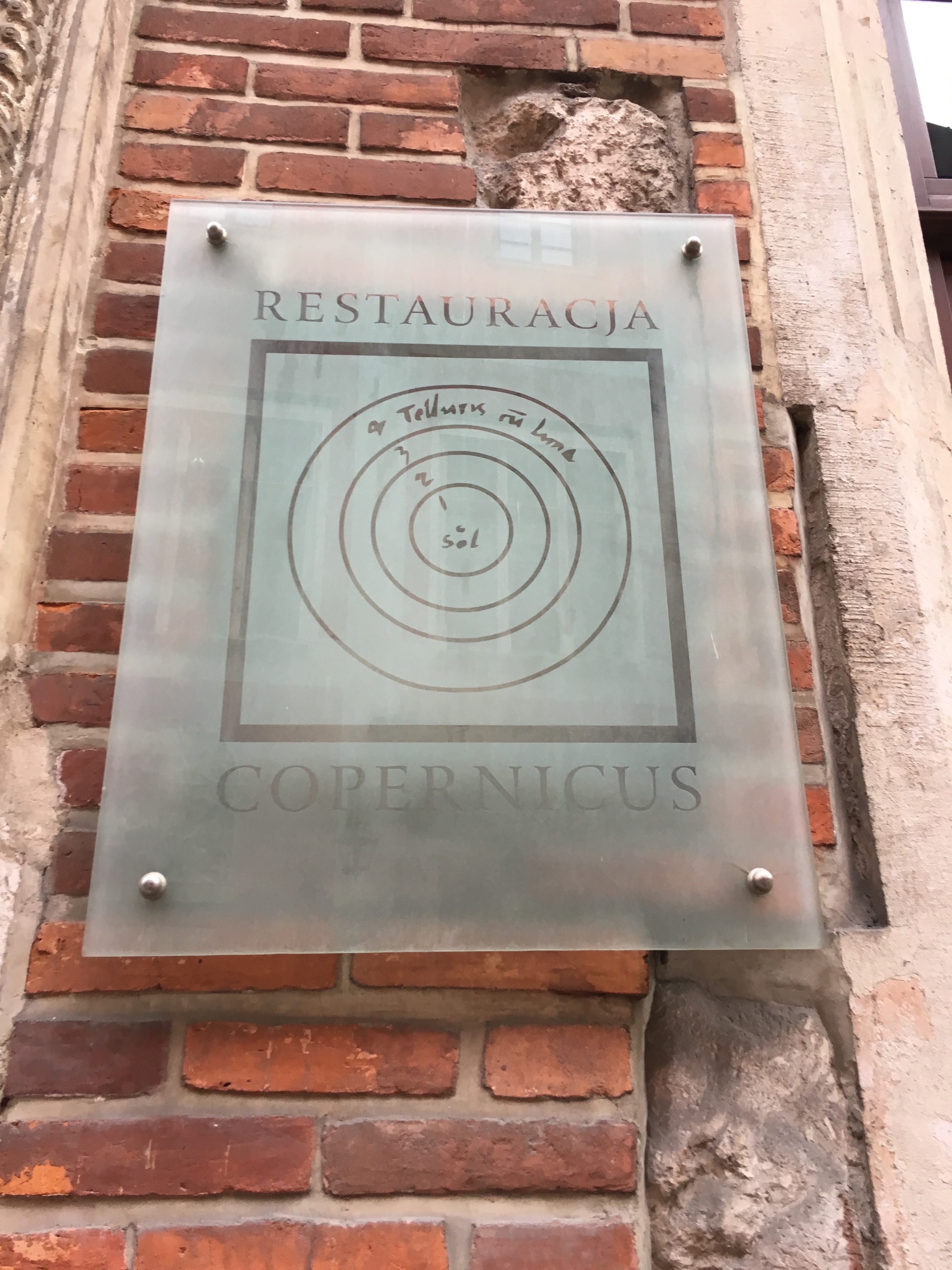}
\end{center}
\caption{Solar system (picture taken by IIB)} 
\label{COPERICUS}
\end{figure}

\item
Just south of the Main Market Square one can see a wonderful connection of Renaissance architecture \& modern sculpture, 
see the {\bf Figure \ref{2arts}} -- if one can find it inside a building.
\begin{figure}[h!]
\begin{center}
\includegraphics[width=5cm]{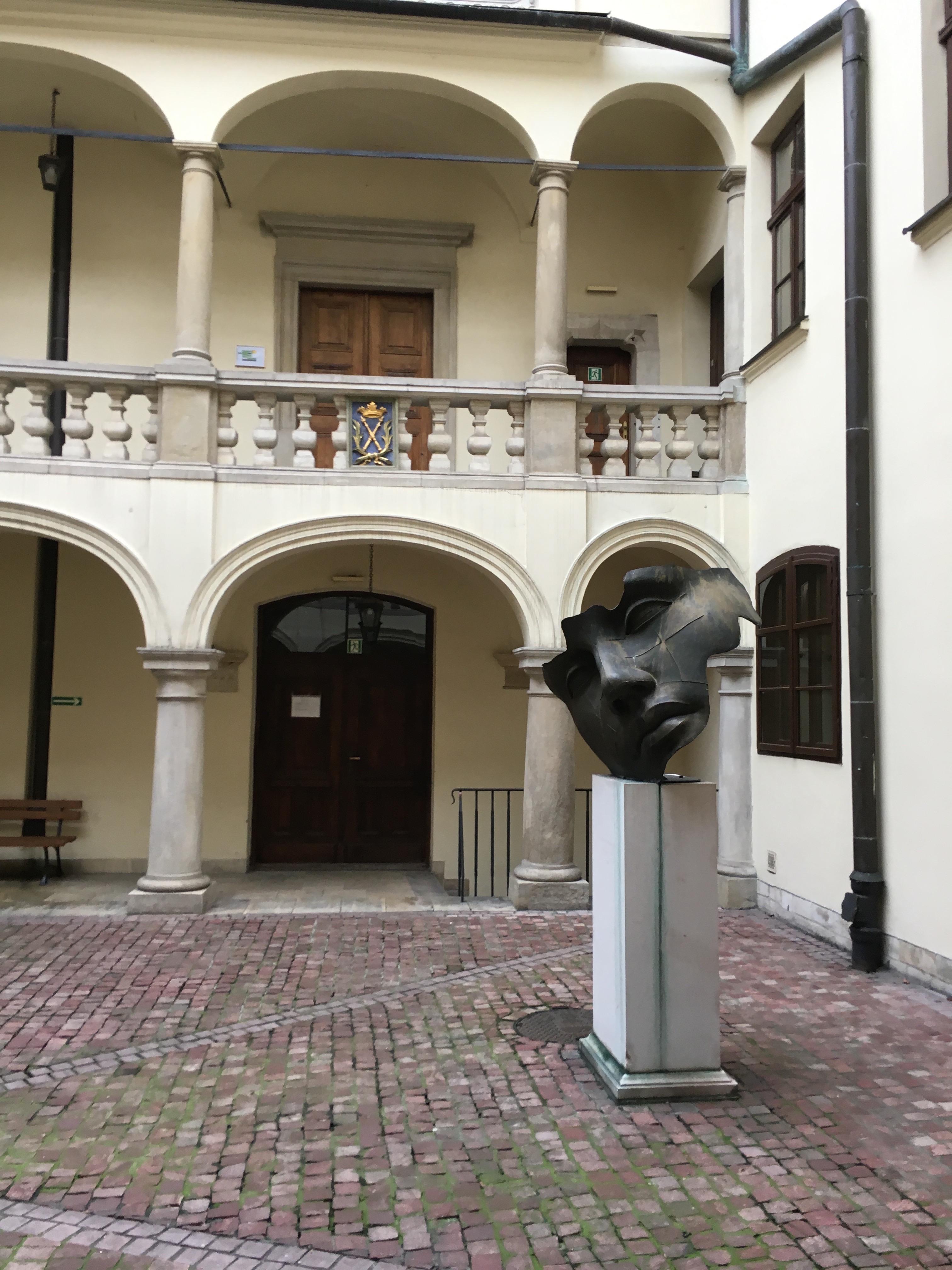}
\end{center}
 \vspace{0cm}
\caption{ Renaissance architecture \& modern sculpture (picture taken by IIB)}
\label{2arts}
\end{figure}

\end{itemize}


\section{Introduction to the `Roads'}

The Greek word `Epiphany' 
means: `manifestation of a divine being an intuitive grasp of reality through something both simple and striking' ! 
I have always been a fan of local Super-Symmetry \& still am; however, 
we are in a different situation: it is {\em neither} simple {\em nor} striking. As I will discuss here in some details: best `fitted' analyses 
of the data  do {\em not} give us the best information about the underlying dynamics -- it is crucial to use correlations with other data \& judgments! 
Furthermore I can admire the courage of the young physicists to deal with the 
challenges in our world on different levels, while listen also to the talks of `mature' colleague like Danish Buras and Swiss Jegerlehner. To make 
progress we have to discuss the disagreements. Fashion does {\em not} help us to go closer to our goals as my Italian colleague Augusto said 
at the conference.

First I will general comments including disagreements I have with some speakers here; some of obvious, while others are more subtle. If a reader 
finds it interesting (I hope), she/he wants details (with many references): N. Uraltsev: "The Heavy Quark Expansion", CRAD96 in Krakow, Acta Phys.Polon.B28(1997)755 
\cite{KOLYA96}; 
N. Uraltsev: "Topics in the Heavy Quark Expansion" in 2000 \cite{KOLYA2000}. Somebody might think it is `old stuff'; however in my view it is still up-to-data of our 
understanding of fundamental forces. 

One might think the choice of words is in the details: HQE vs. HQET. The titles are: HQE = "Heavy Quark Expansion" vs. 
HQET = "Heavy Quark Effective Theory"; in the latter item I want to mention that the applications of HQET in local QCD vs. Lattice QCD 
are different, and I have less problems in the second than the first one. 
The differences go much deeper in their `meaning'. 

The usual HQET papers claim to show the impact of non-perturbative physics:
\beq
{\rm "observable"}\; = \; {\rm perturbative} \; {\rm forces} \;\; \;  + \; \; \; {\rm non-perturbative} \; {\rm forces}
\eeq
Instead Kolya Uraltsev (\& collaborators like Shifman \& me \cite{BSU97}) pointed out that is much deeper to describe the situations by
\beq
{\rm "observable"}\; = \; {\rm short-distance} \; {\rm dynamics} \; \; \; + \; \; \; {\rm long-distance} \; {\rm dynamics}
\eeq
Crucial statements in my view: 
\begin{itemize}
\item
It is not enough to say that OPE is an important theoretical tool: it is the {\em Wilsonian} OPE. The separation of short- vs. long-distances dynamics 
is scale {\em dependent} around 1 GeV for QCD. One might think it is a bad idea and gives more work without better understanding of the underlying dynamics. 
However, I will explain why I disagree with such a `feeling'. 

\item
What the left hand does does not matter what the right hand does? No -- perturbative \& non-perturbative QCD effects have to be treated 
{\em simultaneously} with accuracy; furthermore we have to think about the correlations with experimental  analyses. 

\end{itemize}
These will be discussed with some details or some examples. 

General comments: 
\begin{itemize}
\item
Anomalies -- "deep" or not so far

\item
Wilsonian Operator Product Expansion

\item
Infrared renomalon with non-perturbative QCD

\end{itemize}

Items with some details:
\begin{enumerate}
\item 
Consistent parameterization of the CKM matrix

\item
Definition of quark masses: "${\rm \overline{MS}}$", "kinetic", "PS", `1S', `pole mass'

\item 
$V_{qb}$ [q=c,u]: exclusive vs. inclusive rates and duality

\item 
Broken U- \& V-spin symmetries    

\item
3- \& 4-body final states in beauty \& charm mesons

\item
Challenges for understanding weak decays of beauty \& charm baryons

\item 
The stage of top quarks - in a search for New Dynamics 

\item
Collaboration of HEP \& MEP/Hadrodynamics          
\end{enumerate}
For some of these Sections I have very short comments, while for others I give some discussions with more references. 
In a talk at a conference like this one can only to `paint the landscape', but not beyond. For that one has to go to a 
summer (or winter) schools. 

\section{Anomalies: "deep" or not so}
\label{ANOMALIES}

The word `anomaly' is often used in the literature --  in particular, when one looks for the impact of New Dynamics (ND). 
It is easier to discuss exclusive semi-leptonic transitions. However, the situation is `complex'. 

There is a "quantum anomaly": a classical symmetry is no longer conserved, once one-loop corrections are included. 
In this well-known case of chiral invariance:  for massless quarks we have a "triangle anomaly", since it is produced by a 
diagram with a triangular fermion loop -- or called the "Adler-Bardeen-Bell-Jackiw anomaly":  
\beq
\partial _{\mu}  J_{\mu}^{(5)}= \frac{\alpha_S}{8\pi}\, \tilde G \cdot G (\; + \; m_q\, \bar \psi \gamma _5\psi) \; ; 
\eeq
that is not renormalizable in $4-\epsilon$ dimensions. 
The SM `deals' with that by connecting the world of quarks \& charged leptons (i.e., 3 colors of quarks) 
\footnote{There must be a deep reason for that.}. 

Our community has found `anomalies' in previous \& present data, namely the differences between expectations from the SM vs. measured data as a 
sign of the impact of ND. Even in my view it is not just a fashionable one; we have to work \& think about semi-leptonic transitions in beauty hadrons with several 
examples like $B \to K^* l^+l^-$ \& $\Lambda^0_b \to \Lambda l^+l^-$. One discusses (tiny) rates \& the landscape in $M_{l^+l^-}$. Present data show {\em more} events 
than expected with 3 $\sigma$ uncertainties. Of course, I am not surprised that our colleagues are waiting impatiently to reach 5 $\sigma$ uncertainties or more. 

Allow me to give another lesson in the history: after losing the 1811 battle of Albuera in Spain Marechal Soult said: 
`I had beaten the British -- it was just they did not know when they were beaten.'  He was right on both counts. To `battle with the British' there is an 
analogue to probe the SM \& its limitations: HEP theorists start with a penguin operator $b \to s$ to describe the transitions of $B \to l^+l^- X_s$ as 
$[b\bar q] \to l^+l^- s...\bar q$. In the worlds of hadrons one can measure the final states with $K\pi's$, $2K \bar K$ etc. It makes it in steps: 
$K$, $K^*$, broad resonance $\kappa$, in general $K\pi$'s, $2K \bar K$ etc. The question is: with which certainties can one describe the connection in the 
world of quarks \& gluons with that of hadrons, namely the "duality". I want to pointed out that duality is {\em not} an additional assumption. 
Duality is well-defined in the deep Euclidean region thus avoiding proximity to singularities, cuts induced by hadronic thresholds etc.; then one 
analytically continues it into the Minkowskian domain. There is a price to be paid for this `prize': in general one cannot apply local duality, but averaged 
one over an energy interval of around 1 - 1.5 GeV. Furthermore it is not a mathematical statement: we understand the source of the underlying dynamics; 
it needs some judgment where \& how to apply duality in the world of current quarks \& gluons: 
\beq
[b\bar q] \to l^+l^- s...\bar q \Rightarrow l^+l^- K/K\pi's/2K\bar K...
\eeq
Except that the branch ratios are tiny, the situations are simpler for these transitions: the underlying dynamics can be probed with $M_{l^+l^-}$. 
The situations are much more `complex', when I discuss non-leptonic weak decays below. 
In the future one can probe $b \to d$. The good side is that the SM penguin amplitudes suppressed; unfortunately the landscape has much background.

\section{Wilsonian OPE \& Renormalons}
\label{OPE}

Almost all authors invoke OPE -- but mostly with{\em out} "Wilsonian" prescription. 
One might think it is about bragging right. However, Shifman \& collaborators \cite{SHIFMAN2013} have a long record to emphasize that applying OPE is subtle: 
the Wilsonian OPE has to stop around 1 GeV, not lower.  
It is one thing to draw diagrams, while another thing is understand the underlying dynamics,  
in particular about non-perturbative QCD with some accuracy. I will come back in the next Section about infrared renormalon and later also about the definition of quark masses. 
Mostly I follow the `road' described by Shifman in the Ref.\cite{SHIFMAN2013} with more details now and for the future.

\subsection{First Step to deal with Renormalons}
\label{RENORM}

Dyson pointed out in his famous 1952 paper "Divergences of Perturbation Theory in QED" \cite{DYSON1952} that amplitudes can{\em not} be convergent. 
Later it was realized perturbative series in a QFT are {\em factorially divergent} like $Z=\sum_k C_k \alpha^k k^{b-1}A^{-k}k!$ 
with $k \gg 1$ is the number of loops, $C_k$'s are numerical coefficients of order one, and $b$ \& $A$ are numbers. 
It is traced back to the factorially large number of multi-loop Feynman diagrams. The features responsible for the renormalon factorial divergence is 
the logarithmic running of the effective coupling constant. 

Instead of {\em asymptotic} series one can introduce a Borel transform
\beq
B_Z = \sum_k C_k \alpha^k k^{b-1}A^{-k}  \; ;
\eeq
the singularity of $B_Z (\alpha)$ closest to the origin of the $\alpha$ plain is at a distance $A$, and thus $B_Z (\alpha)$ is {\em convergent}. 
One recovers the original function $Z$ by 
\beq
Z(\alpha) = \int_0^{\infty} dt \; e^{-t}B_Z (\alpha t)
\label{ZALL}
\eeq
The integral representation is well-defined provided that $B_Z(\alpha)$ has {\em no} singularities on the real positive semi-axis in the complex $\alpha$ plane. 
That is not a problem for QED. For other weak couplings it is not trivial, but one can deal with that. 

{\em If} $B_Z(\alpha)$ has a singularity on the real positive semi-axis -- like coefficients $C_k$ are all positive or all negative -- the integrated in the 
Eq.(\ref{ZALL}) become {\em ambiguous}. This ambiguity is of the order of $e^{-A/\alpha}$; more information is needed from the underlying dynamics. 
The question comes from QCD with 
\beq
\alpha_S (Q^2) \simeq \frac{\alpha_S(\mu^2)}{1- \frac{\beta_0 \alpha_S(\mu^2)}{4\pi} {\rm log}(\mu^2/Q^2)} =  
\frac{\alpha_S(\mu^2)}{1+ \frac{\beta_0 \alpha_S(\mu^2)}{4\pi} {\rm log}(Q^2/\mu^2)} \; , \; \beta_0 = 11  - \frac{2}{3}N_f \; ; 
\label{STRONGCOUPL}
\eeq
the energy scale $\mu$ is used to calibrate $\alpha_S (Q^2)$. The good side is: at large scales the strong couplings go down to zero with $Q^2/\mu^2$ (on the log scale) -- 
i.e. "asymptotic freedom". 

On the other hand, there is a true challenge. With $\mu^2 \gg Q^2$  $\alpha_S (Q^2)$ gets larger and larger; thus 
QCD gives us true strong forces at low scales. First one might say it goes to infinite, but that is too naive.  
One has to stop at $\mu \sim 1$ GeV based on perturbative QCD. 

\subsection{Non-perturbative Renormalons}
\label{NONPERT}

It was pointed out first in 1994 that the pole mass is {\em not} well-defined at the non-perturbative level \cite{RENORM1,RENORM2}. 
Furthermore a rather powerful renormalon-based tool was suggested 
for evaluating the corresponding non-perturbative contribution \cite{SHIFMAN2013}. Pole mass is sensitive to large distance dynamics, although this fact is not obvious in 
perturbative calculations. IR contributions lead to an {\em intrinsic uncertainty} in the pole mass of order $\Lambda$ -- i.e., a $\Lambda/m_Q$ power correction. 
It comes from the factorial growth of the high order terms in the $\alpha_S$ expansion corresponding to a singularity residing at the $2\pi/\beta_0$ in the Borel plane. 
Thus one cannot say it is a correction. 

\begin{figure}[h!]
\begin{center}
\includegraphics[width=10cm]{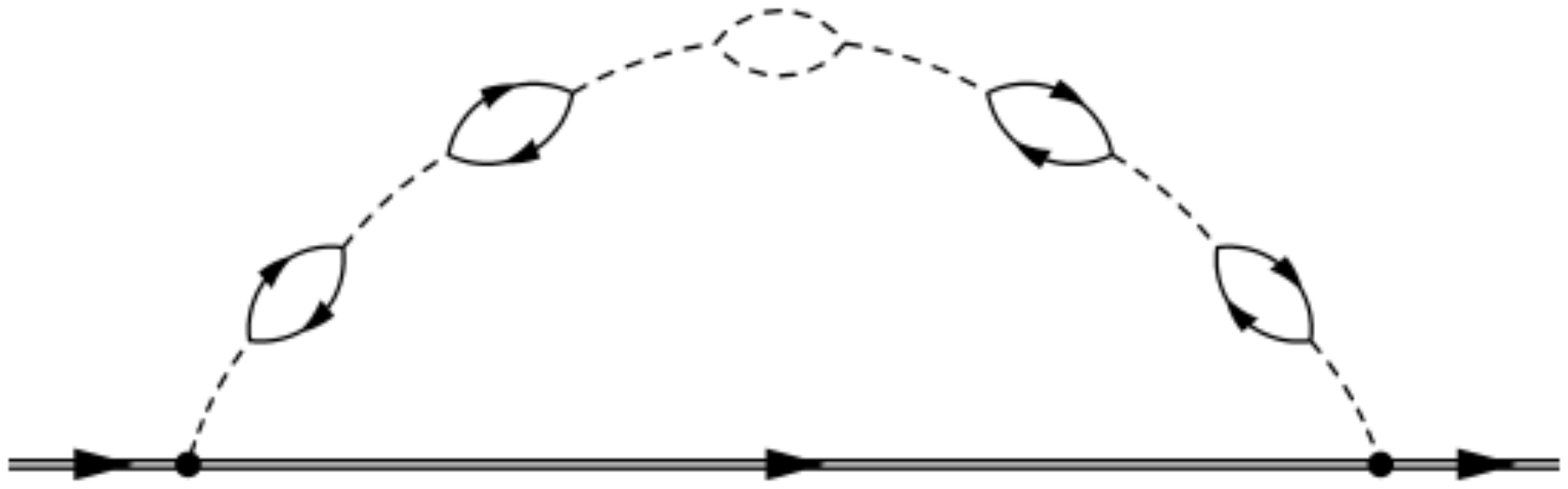}
\end{center}
 \vspace{-11cm}
\caption{ Perturbative diagrams leading to the IR renormalon
uncertainty in $m_Q^{\rm pole}$ of order $\bar \Lambda$.
The number of bubble insertions in the gluon propagator
is arbitrary. The horizontal line at the bottom is the heavy quark Green's function.}
\label{fig5}
\end{figure}

Actually, there are two renormalon-based tools, namely ultraviolate (UV) and infrared (IR) dynamics  
\footnote{"All animals are equal, but some animals are more equal than others!" G. Orwell, `Animal Farm'.}.  One has to include non-perturbative QCD  
with IR one. Those give contribution to $b$ quark mass numerically \cite{BSU97}, see the {\bf Figure \ref{fig5}}:
\bea
\nonumber
m_b^{\rm pole} &=&  m_b( 1 \, {\rm GeV}) +\delta m_{\rm pert}(\leq 1\, {\rm GeV})\simeq \\
&\simeq& 4.55 \, {\rm GeV} + 0.25 \, {\rm GeV} + 0.22 \, {\rm GeV} + 0.38 \, {\rm GeV} + 1\, {\rm GeV} + 3.3\, {\rm GeV}  ... \; ,
\label{MBPOLE}
\eea
where $\delta m_{\rm pert}(\leq 1\, {\rm GeV})$ is the perturbative series taking account of the loop momenta down to zero. 

Top quarks decays before they have produce top hadrons. Still they carry unbroken color symmetry and thus find partners with color 
to produce hadrons with color zero in the FS. I will come back to that below.

\section{Describing the CKM Matrix {\em consistently}}
\label{CKM}

Wolfenstein's parameterization was very smart, easily usable \& well-known. The SM with three families of quarks 
describes the CKM matrix with four parameters, namely $\lambda$, $A$, $\bar \rho$ \& $\bar \eta$. One uses expansion of the  
Cabibbo angle $\lambda = {\rm sin}\theta_C \simeq 0.223$, while $A$, $\bar \rho$ and $\bar \eta$ should be 
of the order of unity \cite{WOLFMAT}. 
It is an important item (in particular about finding the impact of ND), but a subtle one: what does one mean by `maximal' {\bf CP} violation? 
In principle 100 \% asymmetry is possible: I give just three example: $\bar \rho \sim 1$ \& $\bar \eta \sim -1$; $\bar \rho \sim -1$ \& $\bar \eta \sim -0.5$; 
$\bar \rho \sim -0.5$ \& $\bar \eta \sim -0.3$. 

Measured values are $A \simeq 0.82$ as assumed. 
However, {\em measured} $\bar \eta \sim 0.35$ \& $\bar \rho \simeq 0.14$, which are not close to unity; thus we have not real control over systematic uncertainties here. 

The SM produces at least the leading source of {\bf CP} violation in $K_L \to 2 \pi$ and
$B$ decays with good accuracy. Searching for ND we need even precision and to measure
the correlations with other FS's. The landscape of the CKM matrix is more subtle as pointed out
through $ {\cal O}(\lambda ^6)$ consistently  \cite{AHN}: 
\begin{eqnarray}
{\bf {\rm V}}_{\rm CKM} \simeq
\left(\footnotesize
\begin{array}{ccc}
 1 - \frac{\lambda ^2}{2} - \frac{\lambda ^4}{8} - \frac{\lambda ^6}{16}, & \lambda , &
 \bar h\lambda ^4 e^{-i\delta_{\rm QM}} , \\
 &&\\
 - \lambda + \frac{\lambda ^5}{2} f^2,  &
 1 - \frac{\lambda ^2}{2}- \frac{\lambda ^4}{8}(1+ 4f^2)
 -f \bar h \lambda^5e^{i\delta_{\rm QM}}  &
   f \lambda ^2 +  \bar h\lambda ^3 e^{-i\delta_{\rm QM}}   \\
    & +\frac{\lambda^6}{16}(4f^2 - 4 \bar h^2 -1  ) ,& -  \frac{\lambda ^5}{2} \bar h e^{-i\delta_{\rm QM}}, \\
    &&\\
 f \lambda ^3 ,&
 -f \lambda ^2 -  \bar h\lambda ^3 e^{i\delta_{\rm QM}}  &
 1 - \frac{\lambda ^4}{2} f^2 -f \bar h\lambda ^5 e^{-i\delta_{\rm QM}}  \\
 & +  \frac{\lambda ^4}{2} f + \frac{\lambda ^6}{8} f  ,
  &  -  \frac{\lambda ^6}{2}\bar h^2  \\
\end{array}
\right)
\end{eqnarray} 
with $\bar h \simeq 1.35$, $f\simeq 0.75$ \& $\delta _{\rm QM} \sim 90^{o}$ and only
expansion in $\lambda \simeq 0.223$. 
The pattern in flavor dynamics is less obvious for {\bf CP} violation in hadron decays as stated before \cite{BUZIOZ};
the situation has changed: we have to measure the correlations between four triangles, not focus only on the `golden triangle'.
Some of the important points are emphasized: 

\noindent 
(a) The maximal SM value of $S(B^0 \to J/\psi K_S)$ for indirect {\bf CP} violation is $\sim 0.74$. 

\noindent 
(b) For $S(B^0_s \to J/\psi \phi)$ is $\sim 0.03 - 0.05$. 

\noindent 
(c) The SM gives basically zero {\bf CP} value for doubly Cabibbo suppressed transitions. 

\noindent 
One has to measure accurately the correlations with several triangles.

\section{Schemes of Quark Masses} 
\label{QUARKMASSES}

Quark masses are {\em not} observables in general. Therefore I use the word of `Schemes'. 

\subsection{"$\overline{\rm MS}$", "kinetic", "PS"}
\label{GOOD}

$\overline{\rm MS}$ mass $\bar m_Q (m_Q)$ stands for `modified minimal subtraction scheme'. It represents a quantity of computational convenience, in particular when calculating
perturbative contributions in "dimensional regularization"\footnote{It does not necessarily mean we understand the underlying dynamics.}. 
For $\mu  \geq m_Q$ it basically coincides with the running mass in the Lagrangian and is best normalized at
$\mu \sim m_Q$. It is appropriate for describing heavy-flavor {\em production} like $Z^0 \to \bar bb$ and now also $H \to \bar bb$. 
However, it diverges logarithmically for $\mu \to 0$. 

The "kinetic" mass of the heavy quark is regular in the infrared regime including a non-leading source \cite{RENORM2,VOLKOLYA,IBSUV,CICERONE}: 
$\frac{dm_Q^{\rm kin}(\mu)}{d\mu} = - \frac{16}{9}\frac{\alpha_S}{\pi} - \frac{4}{3}\frac{\alpha_S}{\pi}\frac{\mu}{m_Q}  +{\cal O}(\alpha_S^2)$. 
For $b$ quarks $\mu \sim 1$ GeV is the best scale to describe their weak decays 
\footnote{A reader might think, my judgment is `biased'; however, I stay by my statement.}. 
Using $\mu \sim m_b$ instead, it leads to higher-order perturbative corrections that are artificially large, for which one has {\em no} control \cite{IBSUV}.

"PS" = "potential-subtracted": the schemes "kinetic" and "PS" are quite different already on the conception level; technical problems of "PS" arise at 
${\cal O}(\alpha_S^{4})$. Still they are in the same `division' of fundamental physics. I will come back to this point below about top quarks.

\subsection{`Pole mass', `1S'}
\label{NOTGOOD}

A pole mass for quarks is gauge independent and infrared stable in 
perturbative QCD; furthermore it is easy to apply pole mass in Feynman graphs. 
However, it is {\em not} infrared stable non-perturbatively. Make the same statement with different words: 
pole mass depends on long distance dynamics, for what we have little control.

Recent PDG reviews basically ignore the "kinetic" scheme, while focus on the `1S' scheme based on $m_b^{1S} \simeq {\rm M_{\Upsilon(1S)}}/2$ 
\footnote{due to `par ordre du Mufti' (= no right of appeal).}. It claims these schemes give us the same information about underlying dynamics. 
However, it is incorrect, as Uraltsev pointed out \cite{KOLYA2004}: 
$m_b^{1S} = m_b^{\rm pole} [1- C_F^2(\alpha_S^2/8) +{\cal O}(\alpha_S^3, \beta_0\alpha_S^3 {\rm log}\alpha_S ) ] $ -- i.e., also $m_b^{1S}$ is {\em not} well-defined 
at the {\em non}-perturbative level.

\subsection{Short comments}
\label{PINEDA01}

Flavor dynamics is `complex'. At a conference the goal is to `paint' the landscape, but not to discuss the details. However, it is important to give short, but subtle comments. 
I give a reference to an important (\& large) 2001 paper \cite{Pineda2001}. My main disagreements with A. Pineda: his  Abstract does not mention 
some of his important results. However, a careful reader can find it on page 16: 
(a) "... it is achieved by the threshold scheme, i.e. the {\em kinetic}, the PS-like, the 1S ...". I would say the meaning of `threshold' is not obvious. 
When one talks about $b \to c \; [W_{\rm off-shell}^-]$, it means to get one or two charm quarks. 
However, the situations are quite different for $b \to u \; [W_{\rm off-shell}^-]$. 
(b) "Note also that the 1S and PS schemes depend on $\nu_{us}$." At three-loops diagrams 
the ultrasoft scale appears in the static potential and the heavy quarkonium mass. Again, the situations are quite different for the impact of 
perturbative QCD vs. non-perturbative one.

\section{Duality: Measuring $|V_{qb}|$ with $q=c,u$}
\label{DUAL}

The item of "duality" is referred to a very complex situations, namely the connections of the worlds of hadrons vs. quark \& gluons. 
In this Section I give short comments at the very specific case: compare the values of $|V_{cb}|$ and $|V_{ub}|$ from inclusive 
vs. exclusive semi-leptonic amplitudes. 

It seems the difference between the $|V_{cb}|_{\rm incl.}$ vs. $|V_{cb}|_{\rm excl.}$ has become smaller now based on realistic 
theoretical uncertainties, mostly due to LQCD analyses. 

On the other hand, the difference between $|V_{ub}|_{\rm incl.}$ vs. $|V_{ub}|_{\rm excl.}$ has not changed. It has been pointed out that 
the values of $|V_{ub}|_{\rm incl}$ based on the data from $B \to l \nu \pi$'s, while assuming that $B \to l \nu \bar KK ...$ are irrelevant due to a 
traditional understand duality. It is a good assumption -- but {\em local} duality does not work close to thresholds. Maybe the real $|V_{ub}|_{\rm incl}$ 
are smaller and thus solve that challenge. LHCb experiment cannot measure inclusive rates. However, it might go able to go after the rates of 
$B^+ \to l^+\nu K^+K^-$ and $B^0 \to l^+\nu K^+K^-\pi^-$ with non-zero values. Furthermore Belle II should measure values there or limits.

\section{Many-body Final States for $\Delta B \neq 0 \neq \Delta C$ Hadrons}
\label{MANY}

Indirect {\bf CP} violation has been established in $K_L \to2 \pi$ \& $B^0 \to J/\psi K_S$. On the other hand, 
the landscapes are much more `complex' as expected, since direct {\bf CP} asymmetries depend on final state interactions:  
\beq
|T(\bar P \to \bar a)|^2 - |T(P \to  a)|^2 = 4\sum_{a_j,a} T_{a_j,a}^{\rm resc} \; {\rm Im} T^*_a T_{a_j} \; ; 
\eeq 
with{\em out} non-zero re-scattering direct {\bf CP} asymmetries cannot happen, even if there are weak phases 
\cite{1988BOOK,WOLFFSI,FSICP,TIM}. One expects large impact of strong re-scattering, and the LHCb data of suppressed $B \to$ 3 mesons have shown that; 
I will discuss below. It is obvious that the crucial information about the underlying dynamics cannot be found in two-body FS. 
Even so, it is a very good hunting region for the impact of ND, since they can depend only one ND amplitude. 

\subsection{Tools}
\label{TOOL}
One has to think about the tools that can be applied. Not surprisingly, it comes to your mind, namely symmetries broken or not.  
\begin{itemize}
\item
One can apply $SU(3)_{\rm light \; flavor}$ (not $SU(3)_{\rm color}$). The global $SU(3)_{\rm light \; flavor}$ is broken. It was pointed out 
by Lipkin, it helps the thinking by using 3 $SU(2)$: one combines $(u,d)$ quarks for I-spin, while $s \rightleftharpoons d$ for U-spin 
and $s \rightleftharpoons u$ for V-spin symmetries. 

Broken U-spin symmetry with{\em out} V-spin is usable for spectroscopy with a good record. Yet the situation is quite different for weak transitions.  
I give one example from the PDG2017 data {\bf CP} asymmetry: 
\beq
A_{\rm CP} (B^0 \to K^+\pi^-) = - 0.082 {\pm} 0.006
\eeq
(In 1987 Sanda \& I had given a prediction: $A_{\rm CP} (B^0 \to K^+\pi^-) \sim - 0.1$.) 
It shows the impact of Penguin diagrams -- but (semi-)quantitatively.  
Then looks at the PDG2017 data: 
\beq
A_{\rm CP} (B^0_s \to \pi^+K^-) = + 0.26 {\pm} 0.04 \; . 
\eeq
Can we predict this connection?

It had been suggested by Lipkin in 2005 \cite{LIPKIN2005} to use U-spin symmetry 
\footnote{The positive sign in the Eq.(\ref{USPIN}) is not surprizing in the SM.}:
\beq
\Delta = \frac{A_{\rm CP} (B^0 \to K^+\pi^-)}{A_{\rm CP} (B^0_s \to \pi^+K^-)} + \frac{\Gamma (B^0_s \to \pi^+K^-)}{\Gamma (B^0 \to K^+\pi^-)} = 0 \; ; 
\label{USPIN}
\eeq
 The LHCb collab. had published a short 2013 paper \cite{LHCbUSPIN}:
\beq
\Delta_{\rm LHCb} = - 0.02 \pm 0.05 \pm 0.04
\eeq 
saying: "These results allow a stringent test of the validity of the ...". I disagree with this statement for several reasons! First examples from two-body FS: 
\begin{itemize}
\item
Indeed, the value of $\Delta_{\rm LHCb}$ is consistent with zero. 

\item
Yet, it is also consistent with a value $\sim 0.1$ expected for direct {\bf CP} asymmetry for two-body FS.

\item
One has to think about correlations of U-spin symmetry with V-spin one due to re-scattering. 
What about $B^0 \to K^0 \pi^0/K^0\eta$ \& $B^0_s \to \pi^0 K^0 /\eta K^0$? One has to remember that these transitions are affected by oscillations \& indirect 
{\bf CP} violation. 

\item
One can look at the situation with two-body FS of $B^+$ decays: 
\bea
A_{\rm CP}(B^+ \to K_S\pi^+) &=& - 0.017 \pm 0.016    \\
A_{\rm CP}(B^+ \to K^+\pi^0) &=& + 0.037 \pm 0.021  \\
A_{\rm CP}(B^+ \to K^+\eta^{\prime}) &=& + 0.004 \pm 0.011    
\eea
with no sign of {\bf CP} asymmetry, while it was found in 
\beq
A_{\rm CP}(B^+ \to K^+\eta) = - 0.37 \pm 0.08 \; . 
\eeq
It shows the impact of the strong re-scattering. There are two lessons: difference between U- \& V-spin is `fuzzy' due to re-scattering -- and 
we have to go {\em beyond} two-body FS.

\end{itemize}

\item
Probing FS in non-leptonic decays with two hadrons (including narrow resonances) is not trivial to measure {\bf CP} violations, 
if one has enough data for suppressed transitions; theorists can `predict' those \& analyze the data.  On the other hand, one gets 
`just' numbers. We have to remember that two-body FS of of suppressed non-leptonic weak decays are a small part of  charm mesons \& 
tiny ones for beauty mesons; data show that --  it is not surprising.  
Three- \& four-body FS are described by two \& more dimensional plots. There is a price: lots of 
work for experimenters and theorists. There is also a prize: to find the existence of ND and also its features.  

The situations are very different for strange hadrons with $\Delta S = 1\; \& \; 2$ as you listened to my Danish colleague Buras (member of the Bavarian Academy!): 
it is produced by {\em local} operators and FS with only one \& two pions.

\end{itemize}

\subsection{Probing Dalitz plot for $B^{\pm}$}

The data of CKM suppressed $B^+$ decays show no surprising rates for $B^+ \to K^+\pi^-\pi^+$ \& $B^+ \to K^+K^-K^+$ and 
$B^+ \to \pi^+\pi^-\pi^+$ \& $B^+ \to \pi^+K^-K^+$. 

LHCb data from run-1 show averaged direct {\bf CP} asymmetries \cite{LHCb028}:
\bea
\nonumber
\Delta A_{CP}(B^{\pm} \to K^{\pm} \pi^+\pi^-) &=&
+0.032 \pm 0.008_{\rm stat} \pm 0.004_{\rm syst}
\pm 0.007_{\psi K^{\pm}}
\\
\Delta A_{CP}(B^{\pm} \to K^{\pm} K^+K^-) &=&
- 0.043 \pm 0.009_{\rm stat} \pm 0.003_{\rm syst}
\pm 0.007_{\psi K^{\pm}}
\label{SUPP2}
\eea
with 2.8 $\sigma$ \& 3.7 $\sigma$ from zero. Based on our experience with the impact of
penguin diagrams on the best measured $B^0 \to K^+\pi^-$, the sizes of these averaged asymmetries are not surprising; however it does not mean that we
could really predict them. It is very
interesting that they come with opposite signs due to {\bf CPT} invariance.

LHCb data show {\em regional} {\bf CP} asymmetries \cite{LHCb028}:
\bea
\nonumber
A_{CP}(B^{\pm} \to K^{\pm} \pi^+\pi^-)|_{\rm regional} &=& + 0.678 \pm 0.078_{\rm stat}
\pm 0.032_{\rm syst}
\pm 0.007_{\psi K^{\pm}}
\\
A_{CP}(B^{\pm} \to K^{\pm} K^+K^-)|_{\rm regional} &=& - 0.226 \pm 0.020_{\rm stat}
\pm 0.004_{\rm syst} \pm 0.007_{\psi K^{\pm}} \; .
\label{SUPP4}
\eea
"Regional" {\bf CP} asymmetries are defined by the LHCb collaboration: positive asymmetry at low
$m_{\pi ^+\pi ^-}$ just below $m_{\rho^0}$; negative asymmetry both at low and high $m_{K^+K^-}$ values. One should note again the opposite signs in Eqs.(\ref{SUPP4}).
It is not surprising that
"regional" asymmetries are very different from averaged ones. Even when one uses states only from
the SM -- $SU(3)_C \times SU(2)_L \times U(1)$ -- one expects that; it shows the
{\em impact of re-scattering}
due to $SU(3)_C$ (actually $SU(3)_C \times$QED) in general. Of course, our community needs more data,
but that is not enough. There are important questions
and/or statements:
\begin{itemize}
\item
How do we {\em define} regional asymmetries and probe them on the experimental and theoretical sides?

\item
Can it show the impact of broad resonances like $f_0(500)/\sigma$ and $K^*(800)/\kappa$?

\item
Again, the best fitted analyses often do not give us the best understanding of the underlying
fundental dynamics.

\end{itemize}

LHCb data from the run-1 show {\em larger} averaged {\bf CP} asymmetries as discussed above in Eqs.(\ref{SUPP2})
(again with the opposite signs):
\bea
\nonumber
\Delta A_{CP}(B^{\pm} \to \pi^{\pm}  \pi^+\pi^-) &=&
+0.117 \pm 0.021_{\rm stat} \pm 0.009_{\rm syst}
\pm 0.007_{\psi K^{\pm}}
\\
\Delta A_{CP}(B^{\pm} \to \pi^{\pm} K^+K^-) &=&
- 0.141 \pm 0.040_{\rm stat} \pm 0.018_{\rm syst}
\pm 0.007_{\psi K^{\pm}}  \; .
\label{SUPP6}
\eea
It is interesting already with the averaged ones, since $b \Longrightarrow d$ penguin diagrams are more suppressed than $b\Longrightarrow s$ ones.
Again {\bf CP} asymmetries focus on small regions in the Dalitz plots \cite{LHCb028}.
\bea
\nonumber
\Delta A_{CP}(B^{\pm} \to \pi^{\pm} \pi^+\pi^-)|_{\rm regional} &=&
+0.584 \pm 0.082_{\rm stat} \pm 0.027_{\rm syst}
\pm 0.007_{\psi K^{\pm}}
\\
\Delta A_{CP}(B^{\pm} \to \pi^{\pm} K^+K^-)|_{\rm regional} &=&
- 0.648 \pm 0.070_{\rm stat} \pm 0.013_{\rm syst}
\pm 0.007_{\psi K^{\pm}} \; .
\label{SUPP8}
\eea
Again, it should be noted also the signs in Eqs.(\ref{SUPP6}) \& Eqs.(\ref{SUPP8}).
Does it show the impact of
broad scalar resonances like $f_0(500)/\sigma$ and/or $K^*(800)/\kappa$?

First one analyzes the data using model-independent techniques \cite{REIS}, compares them and discuss the results --
but that is not the end of our `traveling'. Well-known tools like dispersion relations are `waiting' to apply -- but we have to do it 
with some `judgement'. I had visited a museum in the north Wall of Krakow and looked at this painting, 
see the {\bf Figure \ref{LADY}}: 
\begin{figure}[h!]
\begin{center}
\includegraphics[scale=0.04]{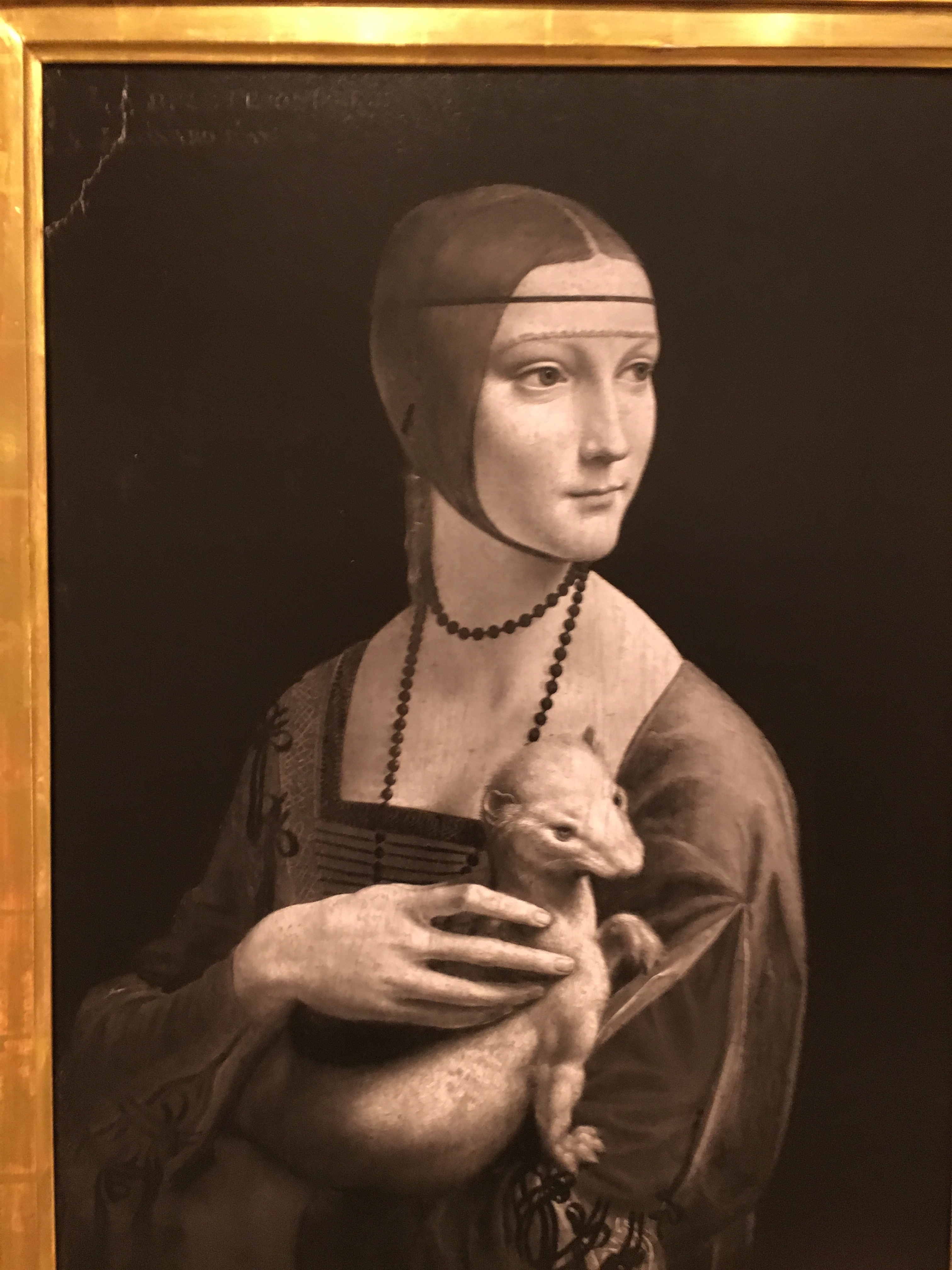}
\end{center}
\caption{`Lady with an Ermine' by Leonardo da Vinci (picture taken by IIB)} 
\label{LADY}
\end{figure}
I was very happy to see it again -- but after more looking at that, I realized I did {\em not} see the `real' paint. It has colors, but {\em pale} ones. 
The real paint with wonderful colors is still in its original part of the museum just a very steps behind this Wall, but it is closed for a year. 
It gives us an idea about the painting of Leonardo da Vinci, but not beyond. 

Coming back to fundamental physics: one has to be prepared for analyses of Dalitz plots (\& beyond); first one has to produce simulations to see both 
the strong \& weak features for hardware \& software of a detector. Yet that is {\em not} the final step. 
The best fitted analyses often do not give us the best information about the underlying dynamics. 
Final steps need judgment based on correlations with other data applying resonances, threshold enhancements etc. with dispersion relations \& other refined tools.


\subsection{{\bf CP} asymmetries in the decays of beauty baryons}

Before I had suggested to probe Dalitz plots of $\Lambda^0_b \to \Lambda \pi^+\pi^-/\Lambda K^+K^-/\Lambda D^-\pi^+$ and  
$\Xi_b^0 \to \Lambda \pi^+\pi^-/\Lambda K^+K^-$ that do not depend on production asymmetries \cite{TIM}. 

However, at the ICHEP2016 conference in Chicago that the LHCb collaboration showed data with evidence for {\bf CP} asymmetry 
in $\Lambda_b^0 \to p \pi^- \pi^+\pi^-$ with a novel idea. It is discussed in \cite{LHCBBARCP} with details. 
In $pp$ collisions one gets different numbers of $\Lambda_b^0$ vs. $\bar \Lambda_b^0$ due to
production asymmetries. Therefore one focuses on {\bf T-}odd moments. LHCb measured the angle between two planes:   
in the rest frame of $\Lambda^0_b$ one plane is defined by $[\vec p \times \vec \pi^-_{\rm fast}]$, while the other one by $[\vec \pi^+ \times \vec \pi^-_{\rm slow}]$; 
likewise for $\bar \Lambda^0_b$. They found evidence for {\bf CP} asymmetry on the level of 3.3 $\sigma$ based in its run-1 of 3 fb$^{-1}$. 
Actually, they  found {\em regional} {\bf CP} asymmetry $\sim 20 \%$ without saying that clearly. In principle it is not surprising due to strong dynamics with 
$\Delta (1232) [\Delta (1600)/\Delta (1620)] \Rightarrow p\pi^-$. We should keep in mind the situations should be affected by different broad resonances, thresholds etc.

Are we lucky to find this effect and its size? Of course, we need more data. Yet, the {\em present} data can give us more information about the underlying dynamics by measuring 
the angle between two different planes: one is defined by $[\vec p \times \vec \pi^-_{\rm slow}]$, while the other one $[\vec \pi^+ \times \vec \pi^-_{\rm fast}]$. Can we 
find {\bf CP} symmetries, too? {\em Regional} ones, where and its size? 

The data are very interesting for several reasons:
 \begin{itemize}
\item
Maybe {\bf CP} asymmetry was found in a decay of a baryon for the first time (except `our existence'); it is for a beauty baryon. 

\item
It is another example that many-body FS are {\em not} a background for the information our community got it from two-body FS.

\item
The plot given at the ICHEP2016 shows the strength of regional {\bf T} asymmetry around 20 $\times 10^{-2}$. Very interesting, but we cannot claim to
understand the underlying dynamics -- yet! Furthermore in the world of quarks \& gluons one looks at
CKM penguin of $b \to d$, where one expects less than for $b \to s$. LHCb data already shown similar lessons
for {\bf CP} asymmetries in $B^+ \to \pi^+\pi^+\pi^-/\pi^+K^+K^-$ vs. $B^+ \to K^+\pi^+\pi^-/K^+K^+K^-$, see 
Eqs.(\ref{SUPP2},\ref{SUPP4},\ref{SUPP6},\ref{SUPP8}) just above.

\item
LHCb collaboration did not get enough data from run-1 to probe $\Lambda_b^0 \to p \pi^-K^+K^-$ \& $\bar \Lambda_b^0 \to \bar p \pi^-K^+K^-$.
It will change very `soon'.

\item
Furthermore,  LHCb collab. can measure rates and {\bf CP}
"regional" asymmetries in $\Lambda_b^0 \to p K^-\pi^+ \pi^-$ and $\Lambda_b^0 \to p K^-K^+ K^-$
`soon' -- and has no competition from other experiments. First we have to discuss
$\Lambda_b^0 \to p \pi^-\pi^+\pi^-$ \& $\Lambda_b^0 \to p \pi^-K^+K^-$ and
$\Lambda_b^0 \to p K^-\pi^+\pi^-$ \& $\Lambda_b^0 \to p K^-K^+K^-$. Will they follow the
same `landscape' for $B^+ \to \pi^+\pi^+\pi^-/\pi^+K^+K^-$ vs.
$B^+ \to K^+\pi^+\pi^-/K^+K^+K^-$ as discussed above qualitatively or not? So say it
with different words: will they show the strengths of `penguin diagrams' in $\Lambda_b^0$ decays or not?
The situations are similar for beauty mesons and beauty baryons or only on the qualitatively way?

\end{itemize}

\section{Top quark in the search for ND}
\label{TOPDYNAMICS}

\begin{figure}[h!]
\begin{center}
\includegraphics[scale=0.40]{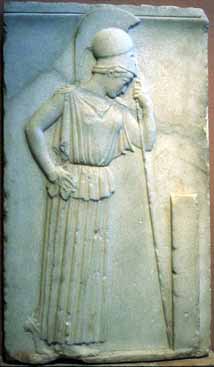}
\end{center}
 \vspace{-0.5cm}
\caption{Greek Goddess Athena} 
\label{ATHENA}
\end{figure}

The landscape of top quark dynamics is very different from $\Delta B \neq 0 \neq \Delta C$, as I  had `painted' it, see the {\bf Figure \ref{LADY}} above. 
To find its direct impact, power is not enough -- we have to think, see the {\bf Figure \ref{ATHENA}}. 

My suggestion is that the 2020 Epiphany Conference in Krakow can mostly focus on describing the landscape of top quark dynamics, namely the productions of a 
pair \& single top quarks and forward vs. center regions of $pp$ collisions; furthermore one discusses {\bf CP}{ asymmetries together $W^{\pm}$ \& $Z^0$ \& $H^0$,  
where one hardly get background from the SM. Finally one discusses the future with new technologies for collisions \& detectors. 
The `future' is defined the time schedule of $\sim$ 30 years, which is beyond my personal `horizon'. I give you only three references coming from the 
10$^{th}$ International Workshop on Top Quark Physics Braga, Portugal, 2017. 
Unfortunately (for me) I did not attend this Workshop. I disagree with several statements given in these papers I found on the internet; 
maybe our real disagreements are smaller, since I am unable to follow the discussions there. It will make progress, but it will need a lot of time.  
My statements below makes my point. 

Now the `top quark community' is hitting the `Systematics Wall' in different ways, see the `Experimental Summary' \cite{TOP2017EXP} 
\footnote{He gave a reference to: V.I. Lenin: `What Is To Be Done?'}: 
\begin{itemize}
\item
take ratios -- go differential\ -- stop and think.

\item
Production rates of $\bar tt$ pairs are powerful handles to constraint the parton distribution functions  (PDF). It has been suggested that $\bar tt$ rates may be the 
relative luminometer of the future for LHC and possible future hadron colliders.

\item
The landscapes for the cross sections of $\bar ttV$ with $V=W,Z$ have changed with the 2016 data, where statistical uncertainties are smaller than the systematic ones; likewise for 
$\bar t t \bar t t$: there are possible hunting regions for ND -- and even more for {\em single} (anti-)top quarks \cite{SINGLETOP}. 

\end{itemize} 
First I make general statements and later examples for special situations. 
There is a comment  about the use of the `words': Nason said in the Abstract in his paper \cite{NASONTOPJAN18}: 
`... calculations interfaced to shower generators (NLO + PS) of increasing accuracy, interfaced to both Pythia8 and Herwig7 Monte Carlo generators.'. 
The first statement: obviously the meaning of his `PS' is quite different from Beneke's `word' as I had discussed in the {\bf Sect.\ref{GOOD}}. 

The second statement is not so short: Nason and I talk about different worlds.

\noindent 
(a) He focus on the perturbative impact of QCD if only with a short comment  claimed `the renormalon ambiguity' is safely below the current 
experimental errors', namely the `ambiguity' of 110 MeV or 250 MeV. I quite disagree. One can{\em not} ignore he works of Shifman \cite{SHIFMAN2013}, which has  
an excellent record \footnote{Of course, I am `biased'.}. There is an important different between perturbative renormalon vs. non-perturbative one. 
Furthermore, how `safe' we are to depend on Monte Carlo generators?  

\noindent 
(b) As I have said above, the `pole mass' is {\em not} well defined. Using simulations \& modeling are one thing (see the {\bf Figure \ref{LADY}} above), while understanding 
the underlying dynamics are quite another thing. Of course, using pole masses are popular -- in particular in experimental papers \& analyses --, but it is only the first step, as I had said above. So far, we are not close to `precision' or even `accuracy'. 

It was said top quarks decay before they can produce top hadrons \cite{ULTRATOP}. Still 
they carry "color" based on a local unbroken QFT; thus they can evolve with other "color" states in connection to produce hadrons without "color" states  in the end. 
It means that the `world' of simulations is {\em less} complex than the FS in the real word.

\subsection{{\bf CP} asymmetries with{\em out} Higgs dynamics}
\label{WITHOUT}

One can measure $pp$ collisions with a pair of $[\bar b W^-]_t[bW^+]_t...$ in the center region with $gg$; or forward(backward) region 
$qg \to q\bar t t$ with $q=u,d$. It is unlikely to find {\bf CP} asymmetries there; on the other hand, we might learn new lessons about very heavy resonances. 

Another `road': $pp \to [\bar bW^-]_t ... [q^{\prime}W^+]_t/  [\bar q^{\prime} W^-]_t ... [bW^+]_t$ with $q^{\prime}=s,d$. One might find 
{\bf CP} asymmetries there; a possible source is an asymmetric Dark Matter. 

One can probe {\bf CP} asymmetry with a {\em single} top quark: $b g \to W^-tg \to W^-[bW^+]_t g$ vs. 
$\bar b g \to W^+\bar t g \to W^+[\bar b W^-]_t g$. Again a possible source is an asymmetric Dark Matter.

\subsection{{\bf CP} asymmetries with {\em on-}shell Higgs dynamics}
\label{WITH}

Collisions at LHC have enough energies to produce very often $pp \to \bar t H^0  t X$; to use different words: one talks about short distance forces like  
$gg\to \bar t H^0 t$. However, can one find these events with a huge background? While I disagree with some statements in these articles, I have to say 
first I admire the courage of these experimenters that enter this challenge.

\section{Summary and a New Alliance for the Future}
\label{ALL}

The ruler of a Greek city in southern Italy once approached the resident sage (Pythagoras) with the request to be educated in mathematics, but in a 
`royal way', since he was busy with many obligations. Whereupon Pythagoras replied with admirable candor: 
`There is no royal way to mathematics.' 

Likewise is there {\em no} `royal insights'  into the inner working of `our' Nature as I try to show first with pictures:  
power is not enough -- we have to think as the {\bf Figure \ref{ATHENA}} shows. 

The painting of Piero della Francesca shows the dream before the crucial battle outside of Rome between Constantine and Maxentius on different dimensions, 
see {\bf Figure \ref{DREAM}} \footnote{Another example of divine manifestation in the old history?}.  
\begin{figure}[h!]
\begin{center}
\includegraphics[scale=1.30]{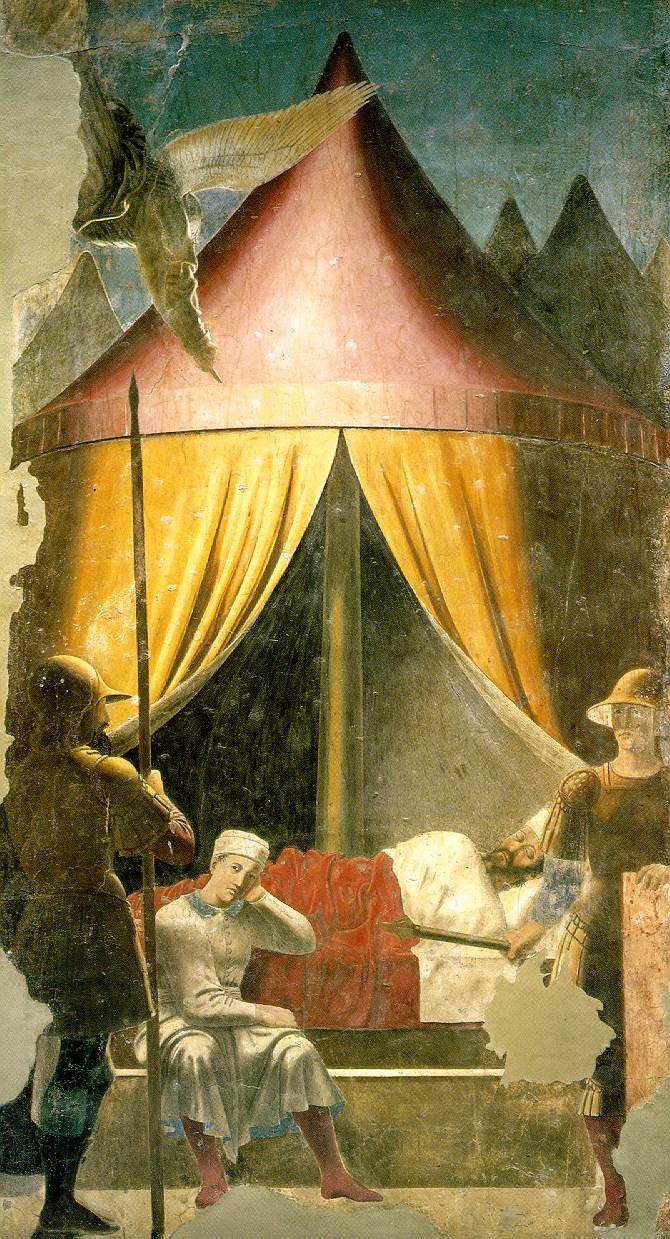}
\end{center}
 \vspace{-0.5cm}
\caption{The dream in the night {\em before} the battle by the Milvian Bridge over the Tiber} 
\label{DREAM}
\end{figure}
Kolya Uraltsev \& I had looked at this painting in person and realized that it is symbol of a true collaboration.

Our community proceeds in steps: first one uses models to describe the data and then model-independent analyses. 
However, those should not be the final step(s). 
Often best fitted analyses do not give us the best information about the underlying dynamics. How to do that? 
We have theoretical tools with a good record like dispersion relations \& other refined tools. 
They are `waiting' -- it `only' needs to work with {\em judgements} and tests it with correlations with other data! 
Yes, the data are the referees, but in the end -- theorists should not be the slaves of the data.

In the previous century we had talked about fundamental physics: Nuclear Physics at low energies, while HEP at high energies;  
flavor dynamics are part of HEP. 
In this century one thinks (or should) about Nuclear Physics and MEP and HEP. Probing jets, Higgs \& top quarks dynamics 
and direct SUSY is the `job' for HEP still again. However, the landscape is more complex with many interconnected parts: decays of strange/beauty/charm 
hadrons, where tools applied to Dalitz plots with dispersion relations etc. We have to go for accuracy and even precision to find the impact of ND. 
To make progress, it is crucial to connect the world of hadrons, where MEP applies --  or with a better choice of word, namely "hadro-dynamics" --  
with the world of quarks \& gluons, where HEP works; it is highly non-trivial.

\section{Personal Epilogue from my week in Krakow}

In a museum of Krakow that is inside of the north Wall of the old center I have seen a very good Roman sculpture to show the goddess `Minerva/Athena'.  
I saw a group of pairs of ladies, where one was blind and the other was a guide: the blind one was allowed to {\em touch} this sculpture in some details --  
a wonderful experience!  

\vspace{0.3cm}

{\bf Acknowledgments:} This work was supported by the NSF under the grant number PHY-1520966. I truly thank the organizers of the 2018 Epiphany Conference! 
If a reader of my article has seen the center of Krakow, she/he should understand why I used pictures, not only diagrams. 
\vspace{2mm}



\end{document}